\documentclass[12pt]{iopart}
\usepackage{iopams}

 \newtheorem{theorem}{Theorem}
\newtheorem{lemma}{Lemma}

\newcommand{\proof } {\noindent{\bf{Proof.}} }

\begin{document}
\title{Analysis of a convenient information bound for general quantum channels}

\author{C J O'Loan}

\address{School of Mathematics and Statistics, University of St Andrews, 
 KY16~9SS, UK}

\ead{cjo2@st-and.ac.uk}

\begin{abstract}
Open questions from Sarovar and Milburn (2006 {\em J.Phys. A: Math. Gen.} {\bf 39} 8487) are answered. Sarovar and Milburn derived a convenient upper bound for the Fisher information of a one-parameter quantum channel.  
They showed that for quasi-classical models their bound is achievable and they gave a necessary and sufficient condition for positive operator-valued measures (POVMs) attaining this bound.
They asked (i) whether their bound is attainable more generally, (ii) whether explicit expressions for optimal POVMs can be derived from the attainability condition.
We show that the symmetric logarithmic derivative (SLD) quantum information is less than or equal to the SM bound, i.e.\ $H(\theta) \leq C_{\Upsilon}(\theta)$ and we find conditions for equality. As the Fisher information is less than or equal to the SLD quantum information, i.e.\ $F_M(\theta) \leq H(\theta)$, we can deduce when equality holds in $F_M(\theta) \leq C_{\Upsilon}(\theta)$. Equality does not hold for all channels. As a consequence, the attainability condition cannot be used to test for optimal POVMs for all channels. These  results are extended to multi-parameter channels. 
\end{abstract}

\section{Introduction} \label{sec:intro}
This paper looks at the distinguishability of quantum channels and answers open questions from Sarovar and Milburn \cite{saromilb06}.
A quantum channel is a trace-preserving completely positive map sending density matrices to density matrices. 
Any quantum channel can be represented using Kraus operators  [2,3]  $E_{k}$ as
\begin{eqnarray*}
 \rho_{0} \mapsto  \sum_{k} E_{k} \rho_{0}E_{k}^{\dagger}
\label{eq:0pchannel}
\end{eqnarray*}
where
\begin{eqnarray*}
\sum_{k} E_{k}^{\dagger}E_{k} = \mathbb{I}.
\label{eq:0pchannel}
\end{eqnarray*}
In the case where there is only one $E_k$ the channel is unitary. In this paper we look at the case where the input and output states live in identical Hilbert spaces, i.e.   $\mathcal{H}_1 = \mathcal{H}_2 = \mathbb{C}^d$. Most quantum information processes can be represented as quantum channels. In practice, quantum channels are not known {\em a priori} and the estimation of them is of great importance.

There are several ways to estimate a quantum channel. One approach is quantum process tomography, which is discussed in chapter $8$ of \cite{chuang00}. One needs to estimate how the channel acts on different bases of the Hilbert space plus linear combinations thereof.  A problem with this method is that in many practical situations it is not possible to prepare these input states in the laboratory \cite{demartinietal03}.

Another approach is to assume that the channel comes from a parametric family of channels [5-10].
This is a reasonable assumption. A family of channels parametrized by a real parameter $\theta$ can be represented by Kraus operators depending on $\theta$ as
\begin{eqnarray}
 \rho_{0} \mapsto  \sum_{k}E_{k}(\theta) \rho_{0} E_{k}^{\dagger}(\theta)  .
\label{eq:1pchannel}
\end{eqnarray}
We choose a suitable input state $\rho_0$ such that the output state is in one-to-one correspondence with the channel. Since a specific value of $\theta$ corresponds to a specific channel, estimation of the channel reduces to a parameter estimation problem. 

 We will now look briefly at the theory of one parameter quantum estimation. Quantum estimation is concerned with estimating (especially optimally) quantum states and processes. Two important mathematical objects in quantum estimation are density matrices and  positive operator-valued measures (POVMs). A density matrix represents the state of the quantum system. The density matrix of a $d$-dimensional state is a $d \times d$ non-negative, Hermitian matrix, with trace $1$, i.e.
 \begin{eqnarray*}
 \langle v | \rho | v \rangle \geq 0,\quad  \forall | v \rangle, \quad \rho^{\dagger} = \rho , \quad \tr\{\rho\} =1.
 \end{eqnarray*}
 A POVM is represented by a set of operators $\{ M_m \}$ which are Hermitian, non-negative and sum to the identity, i.e.
  \begin{eqnarray*}
M_m^{\dagger} =  M_m , \quad \langle v | M_m | v \rangle \geq 0,\quad  \forall | v \rangle,  \quad \sum_m M_m = \mathbb{I}.
 \end{eqnarray*}
 Given a state $\rho(\theta)$ with an unknown parameter $\theta$ and a POVM $\{ M_m \}$, the probability density of the measurement yielding the result $x_m$ is
 \begin{eqnarray*}
p(x_m ; \theta) = \tr \{  \rho(\theta)  M_m \} .
\label{eq:born}
\end{eqnarray*}
These probabilities usually depend on $\theta$.   In practice, we repeat a measurement $N$ times.  The outcomes of the measurement depend probabilistically on the parameter $\theta$. We then choose an estimator which gives us an estimate of $\theta$ from the measurement results. 
For optimal estimation of a state, we have to choose the POVM and estimator that give us the most `information' about the state. 
In the case of quantum channels, we have to optimize further over input states.
 For some channels there exist input states which do not change under the action of a channel.  Hence, we try to choose an input state that gives us the most `information' about the channel.

A standard way of quantifying the performances of input states and POVMs is to use Fisher information.  Fisher information tells us the amount of `information' about $\theta$ contained in a measurement result. Fisher information is defined as
\begin{eqnarray*}
 F_{M}(\theta) &\equiv&  \int d\xi  p(\xi;\theta) \left( \frac{d \ln p(\xi;\theta)}{d \theta} \right)^2  \\
&=&  \int d\xi \frac{1}{p(\xi;\theta)} \left( \frac{d p(\xi;\theta)}{d \theta} \right)^2.
\label{eq:FI00}
\end{eqnarray*}
If the measurement outcomes are discrete with probabilities $p_1(\theta),\dots ,p_n(\theta)$, then the Fisher information can be expressed as
\begin{eqnarray*}
 F_{M}(\theta) = \sum_{k=1}^{n} \frac{1}{p_k (\theta)} \left( \frac{d p_k (\theta)}{d \theta}\right)^2 .
\label{eq:FI01}
\end{eqnarray*}
The importance of Fisher information is seen in the Cram\'er--Rao inequality. This states that the variance of an unbiased estimator $t(x)$ is greater than or equal to the reciprocal of the Fisher information, i.e.
\begin{equation}
\mathrm{var}_{\theta}[t(x)] \geq \frac{1}{ F_{M}(\theta)}.
\label{eq:CRineq}
\end{equation}
Under mild regularity conditions for $p(x ; \theta)$, a maximum likelihood estimator achieves this lower bound asymptotically \cite{Nahi69}. The larger the Fisher information, the more accurately we can estimate the unknown parameter. A standard approach to estimation is to choose the procedure  which maximizes the Fisher information and use the maximum-likelihood estimator.

An important quantity in quantum estimation theory is the symmetric logarithmic derivative (SLD) quantum information. The SLD quantum information $ H(\theta)$ is an upper bound on the Fisher information \cite{braunsteincaves94}, i.e.
\begin{eqnarray}
F_{M} (\theta) \leq H(\theta) .
 \label{eq:smbound2}
 \end{eqnarray}
The SLD quantum information gives the maximal Fisher information in a measurement on a quantum system and is always achievable, at least locally, for one-parameter models\cite{braunsteincaves94}. Putting together (\ref{eq:CRineq}) and (\ref{eq:smbound2}) we get the quantum Cram\'er--Rao inequality 
\begin{equation}
\mathrm{var}_{\theta}[t(x)] \geq \frac{1}{ H(\theta)}.
\label{eq:QCRineq}
\end{equation}
Optimal estimation of one-parameter quantum channels based on the SLD involves: first, choosing an input state which maximizes the SLD quantum information of the output state; second, choosing a measurement which maximizes the Fisher information, i.e.\ gives equality in (\ref{eq:smbound2}).

For a one-parameter family of states $\rho(\theta)$, the SLD quantum information is defined as
\begin{equation*}
H(\theta) = \tr\{\rho(\theta)\lambda(\theta)^2\},
 \label{qi6}
\end{equation*} 
where the SLD quantum score $\lambda(\theta)$ is a self-adjoint solution of
\begin{equation}
\frac{d \rho(\theta)}{d\theta} = \frac{1}{2}\left\{ \rho(\theta)\lambda(\theta)+\lambda(\theta)\rho(\theta) \right\}.
 \label{qi5}
\end{equation}
To achieve equality in (\ref{eq:smbound2}) we have to choose a POVM $\{ M_m \}$ satisfying \cite{barndorffnielsengill00}
\begin{equation}
M_{m}^{1/2} \lambda(\theta) \rho(\theta)^{1/2} = \xi_m(\theta)  M_{m}^{1/2} \rho(\theta)^{1/2}, \quad  \qquad \forall m,
\label{eq:eupsrho1}
\end{equation}
where $\xi_m(\theta)$ is a real number. Since $\lambda(\theta)$ is self-adjoint, it can be diagonalized, i.e.\ we can write it in the form
\begin{equation*}
\lambda(\theta) = \sum_i c_i(\theta) | e_i (\theta) \rangle \langle e_i (\theta) |, \quad \langle e_i (\theta) | e_j (\theta) \rangle = \delta_{ij}.
\end{equation*}
The POVM $\{  | e_m (\theta) \rangle \langle e_m (\theta) | \}$ satisfies (\ref{eq:eupsrho1}) and so gives equality in (\ref{eq:smbound2}) \cite{braunsteincaves94}.  This POVM will generally depend on the unknown parameter $\theta$, and so we have to use an adaptive approach \cite{barndorffnielsengill00}. Usually we will have $N$ copies of the state we wish to estimate.  The adaptive measurement involves measuring a small number $n$ of copies to get a rough estimate $\hat \theta$ of the parameter. Then we use the optimal POVM for $\rho(\hat \theta)$ on the remaining $N - n$ copies.

\subsection{Sarovar and Milburn's approach} \label{sec:intro}
Sarovar and Milburn focussed on the problem of estimation of a one-parameter quantum channel when the input state is fixed. They looked at finding measurements maximizing the Fisher information. One problem with the SLD quantum information is that it is often very cumbersome to compute. They looked for an achievable upper bound on the Fisher information which is easy to compute. For one-parameter channels of the form (\ref{eq:1pchannel}), where $\rho_{0}$ is a completely known pure input state, Sarovar and Milburn \cite{saromilb06} found a simply computable upper bound (given in (\ref{eq:smboundups}) below) on the Fisher information. They also gave a necessary and sufficient condition for a POVM to achieve this bound. They derived the inequality
 \begin{eqnarray}
F_{M} (\theta) \leq C_{E}(\theta) , \qquad \forall E,
 \label{eq:smbound}
 \end{eqnarray}
where $E$ corresponds to a specific set of Kraus operators $\{ E_k \}$ and 
\begin{eqnarray}
C_{E}(\theta)  = 4 \sum_{k} \tr \{   E_{k}'(\theta)  \rho_{0} E_{k}'^{\dagger}(\theta) \}, \qquad E_{k}'(\theta) = \frac{d }{d \theta}E_{k}(\theta).
\label{ce}
 \end{eqnarray}
The quantity $C_{E}(\theta)$  depends on the Kraus representation \cite{saromilb06}. To obtain a bound which depends only on the channel and not on the Kraus representation, we choose the bound given by the canonical Kraus decomposition. The canonical Kraus operators  $\{ \Upsilon_{k}(\theta) \}$ are uniquely defined as the set of Kraus operators satisfying
\begin{eqnarray}
\tr \{ \Upsilon_{k}(\theta)  \rho_{0}  \Upsilon_{j}^{\dagger }(\theta)  \} = \delta_{jk} p_{k}(\theta), \qquad \forall j,k.
\label{eq:condkco}
\end{eqnarray}
From (\ref{eq:smbound}) we get 
\begin{eqnarray}
F_{M} (\theta) \leq C_{\Upsilon}(\theta). 
 \label{eq:smboundups}
 \end{eqnarray}
Throughout the rest of this paper we will refer to (\ref{eq:smboundups}) as the SM bound. If this bound is not uniformly attainable, i.e.\ there does not exist some POVM $\{ M_m \}$ such that $F_{M} (\theta) = C_{\Upsilon}(\theta)$ for all $\theta$, then no bound of the form (\ref{ce}) is uniformly attainable \cite{saromilb06}. To achieve equality in (\ref{eq:smboundups}) the POVM $\{ M_m \}$ must satisfy
\begin{equation}
M_{m}^{1/2}  \Upsilon_{k}'(\theta) \rho_{0}^{1/2} = \xi_{m}(\theta)  M_{m}^{1/2}\Upsilon_{k}(\theta)  \rho_{0}^{1/2}, \qquad \forall m, k,
\label{eq:eupsrho}
\end{equation}
for some real $\xi_m(\theta)$. For channels with quasi-classical output states, it was shown in \cite{saromilb06} that this bound is attainable. Quasi-classical models are parametrized families of states  for which only the eigenvalues (but not the eigenvectors) depend on $\theta$. These models have spectral decomposition
 \begin{eqnarray*}
\rho_{out}(\theta) = \sum_k p_k(\theta) | w_k \rangle \langle w_k |,
\label{eq:quasic}
\end{eqnarray*}
 with fixed eigenvectors $ | w_k \rangle$ that do not depend on $\theta$. 
 
Sarovar and Milburn asked (i) whether their bound (\ref{eq:smboundups}) is attainable more generally,  (ii) whether explicit expressions for optimal POVMs can be derived from the attainability conditions (\ref{eq:eupsrho}). It is very important for an upper bound on Fisher information to be attainable, otherwise it gives us an unrealistic view of how well we can estimate a parameter. In this paper, we give the conditions for which  (\ref{eq:smboundups}) is attainable.  It is also shown that condition (\ref{eq:eupsrho}) cannot be used generally to test for optimal POVMs. 

\section{One-parameter channels}
In this section we answer questions from \cite{saromilb06} about the SM bound for one-parameter channels.  When the input state is pure with $\rho_{0} = | \psi_{0} \rangle \langle \psi_{0} |$, condition (\ref{eq:condkco}) for the canonical Kraus decomposition is equivalent to the condition 
\begin{eqnarray*}
\langle v_{j}(\theta)  | v_{k}(\theta) \rangle  = \delta_{jk} p_{k}(\theta), \qquad  \mathrm{where}   \quad  | v_{k}(\theta)  \rangle  =  \Upsilon_{k} (\theta) | \psi_{0}  \rangle .
\end{eqnarray*}
The output state is 
\begin{eqnarray*}
\rho_{out}(\theta) = \sum_{k} | v_{k}(\theta)  \rangle  \langle v_{k}(\theta) |.
\end{eqnarray*}
This can be rewritten as
\begin{eqnarray}
\rho_{out}(\theta) = \sum_{k} p_{k}(\theta) | w_{k}(\theta) \rangle  \langle w_{k}(\theta) | , \qquad | w_{k}(\theta)   \rangle  =\frac{1}{\sqrt{p_{k}(\theta)}}| v_{k}(\theta)  \rangle.
\label{eq:outstate}
\end{eqnarray}
Thus the canonical decomposition leads to the spectral decomposition of the output state \cite{saromilb06}. The SLD quantum information, $H(\theta)$, and the SM bound, $C_{\Upsilon}(\theta)$, can be expressed as
(see Appendices A and B)

\begin{eqnarray}
H(\theta) &=& \sum_{k, p_k \neq 0} \frac{p_{k} '^{2}}{p_{k}} +   \sum_{  j < k, p_j + p_k > 0  } 4\frac{ ( p_{j} - p_{k})^2}{p_{j} + p_{k}} |\langle w_{j}' | w_{k}  \rangle|^2   ,
\label{eq:Hsldr}\\
 C_{\Upsilon}(\theta) &=& \sum_{k, p_k \neq 0} \frac{p_{k}'^{2}}{p_{k}} +  \sum_{ j < k, p_j + p_k > 0 } 4 (p_{j} + p_{k}) |\langle w_{j}' | w_{k}  \rangle |^2  \nonumber \\
 &&   + 4    \sum_{k, p_k \neq 0} p_{k} |\langle w_{k}' | w_{k}  \rangle |^2 .
\label{eq:samr}
\end{eqnarray}

{\noindent{\bf{Remark 1}} }
It follows from (\ref{eq:outstate}) and (\ref{eq:samr}) that the SM bound can be described solely in terms of the family of output states.

{\noindent{\bf{Remark 2}} }
The SM bound was originally derived as an upper bound on the Fisher information for a one-parameter quantum channel. Because all quantum states and families of quantum states are Hermitian, they can be diagonalized, i.e. written in the form
\begin{eqnarray*}
\rho(\theta) = \sum_k p_{k}(\theta) | w_{k}(\theta) \rangle  \langle w_{k}(\theta) |.
\label{eq:outstate2}
\end{eqnarray*}
By re-writing the SM bound as (\ref{eq:samr}), we can extend it to an upper bound on the Fisher information for one-parameter families of states.

\begin{lemma}
\begin{eqnarray}
 H(\theta) \leq C_{\Upsilon}(\theta),
\label{eq:ineqcs2}
\end{eqnarray}
i.e.\ the SLD quantum information is less than or equal to the SM bound. 
\label{eq:ineqcs}
\end{lemma}

\proof
The first terms in (\ref{eq:Hsldr}) and (\ref{eq:samr}) are identical. 
\begin{equation*}
C_{\Upsilon}(\theta) - H(\theta) =  A_{C} - A_H +  B_{C}
\end{equation*}
where 
\begin{eqnarray*}
A_{H} &=&  \sum_{  j < k, p_j + p_k > 0 } 4\frac{( p_{j} - p_{k})^2}{p_{j} + p_{k}} |\langle w_{j}' | w_{k}  \rangle|^2, \\
A_{C} &=&  \sum_{  j < k, p_j + p_k > 0 } 4(p_{j} + p_{k}) |\langle w_{j}' | w_{k}  \rangle |^2 ,\\
B_{C} &=& 4 \sum_{k, p_k \neq 0} p_{k} |\langle w_{k}' | w_{k}  \rangle |^2 .
\label{eq:equality1}
\end{eqnarray*}
The terms $A_C$ and $A_H$ are symmetric in $j$ and $k$ due to (\ref{eq:diffwjwk}). Now
\begin{eqnarray*}
A_C - A_H &=& 2 \sum_{  j \neq k, p_j + p_k > 0 } \frac{( p_{j} + p_{k})^2 - ( p_{j} - p_{k})^2}{p_{j} + p_{k}} |\langle w_{j}' | w_{k}  \rangle|^2, \\
&=& 8 \sum_{  j \neq k, p_j + p_k > 0 } \frac{p_{j} p_{k}}{p_{j} + p_{k}} |\langle w_{j}' | w_{k}  \rangle|^2.
\end{eqnarray*}
We can change the range of the summation  to $j \neq k$ where $p_j,p_k > 0$. Adding $B_C$ we get
\begin{equation}
C_{\Upsilon}(\theta) - H(\theta) = 8 \sum_{  j, k, p_j ,p_k > 0 } \frac{p_{j} p_{k}}{p_{j} + p_{k}} |\langle w_{j}' | w_{k}  \rangle|^2.
\label{cmh}
\end{equation}
The right hand side of (\ref{cmh}) is always greater than or equal to zero, we have (\ref{eq:ineqcs2}).
 \begin{lemma}
Equality holds in (\ref{eq:ineqcs2}) if and only if the channels satisfy
\begin{eqnarray}
\langle w_{j}' | w_{k}  \rangle = 0, \quad \forall j,k, \quad \mathrm{with} \quad  p_j, p_k > 0.
\label{wjwk0}
\end{eqnarray}
  \end{lemma}
  \proof
This follows directly from (\ref{cmh}).
\label{lemmylem}

  \begin{lemma}
For channels for which $p_j(\theta) > 0$ for all $j$ and $\theta$, the bound (\ref{eq:ineqcs2}) is achievable if and only if the channel is quasi-classical.
\label{lemc}
\end{lemma}
\proof 
Equality holds in (\ref{eq:ineqcs2}) if and only if (\ref{wjwk0}) is satisfied. For (\ref{wjwk0}) to be satisfied when $p_j(\theta) > 0$ for all $j$ and $\theta$, we require that  $|w_{j}'  \rangle$ has zero components along every vector $|w_{k} \rangle$. This is possible only if $|w_{j}'  \rangle = 0$ and hence the channel is quasi-classical.

\begin{lemma}
For unitary channels the bound (\ref{eq:ineqcs2}) is achievable if and only if
\begin{equation}
\tr \{ U(\theta) \rho_{0} U(\theta)'^{\dagger}  \} = 0.
\label{eq:unitconds}
\end{equation}
\label{lemd}
\end{lemma}
\proof 
Equality holds in  (\ref{eq:ineqcs2}) if and only if (\ref{wjwk0}) is satisfied. For unitary channels there is only one non-zero $p_j$ and $| w_j \rangle = U(\theta) | \psi_0 \rangle$, where $\rho_0 =  | \psi_0 \rangle \langle \psi_0 |$.  For (\ref{wjwk0}) to be satisfied, we require $\langle w_j ' | w_j \rangle = 0$. This is equivalent to (\ref{eq:unitconds}).

{\noindent{\bf{Example 1}} }
There exist channels which are neither quasi-classical or unitary for which equality holds in (\ref{eq:ineqcs2}). The channel with output states
\begin{equation*}
\rho_{out}(t) = t^2 | v_1(t) \rangle \langle v_1(t)  | + ( 1 - t^2 ) | v_2(t)  \rangle \langle v_2 (t) |, \quad 0 < t < 1,\\
\end{equation*}
where
\begin{equation*}
| v_1(t)  \rangle = ( t, \sqrt{ 1 - t^2}, 0 )^T, \quad  | v_2(t)  \rangle = ( 0, 0, 1 )^T,
\end{equation*}
satisfies (\ref{wjwk0}), and so equality holds in (\ref{eq:ineqcs2}).
 
  \begin{theorem}
 The SM bound (\ref{eq:smboundups}) is achievable only for channels satisfying 
 \begin{eqnarray*}
\langle w_{j}' | w_{k}  \rangle = 0, \quad \forall j,k, \quad \mathrm{with} \quad  p_j, p_k > 0.
\end{eqnarray*}
   \end{theorem}
 \proof
The SM bound (\ref{eq:smboundups}) follows from (\ref{eq:smbound2}) and (\ref{eq:ineqcs2}). The SM bound is attainable  only when there is equality in  (\ref{eq:smbound2}) and (\ref{eq:ineqcs2}). It is always possible to find a POVM $M_{\theta}$, depending on $\theta$, which achieves equality in (\ref{eq:smbound2}) for one-parameter channels \cite{braunsteincaves94}. However, equality holds in  (\ref{eq:ineqcs2}) only for channels satisfying (\ref{wjwk0}).

 \begin{lemma}
The SLD quantum information is less than or equal to all $C_{E}(\theta)$ of the form (\ref{ce}), i.e.
 \begin{equation}
 H(\theta) \leq C_{E}(\theta).
   \label{hlessce2}
 \end{equation}
  \label{hlessce}
   \end{lemma}
  \proof
  We split this proof up into two cases:
 \begin{enumerate} 
\item[(i)] When (\ref{eq:smbound}) is attainable, (\ref{eq:smboundups}) is also attainable \cite{saromilb06}. In this case  $C_{\Upsilon}(\theta) \leq C_{E}(\theta)$ for all other Kraus decompositions \cite{saromilb06}.  Using (\ref{eq:ineqcs2}) we get $H(\theta) \leq C_{E}(\theta)$.

\item[(ii)] When (\ref{eq:smbound}) is not attainable then $F_{M}(\theta) <  C_{E}(\theta)$ for all $M$.
 For one-parameter channels there always exists a measurement $M_{\theta}$ such that $F_{M_{\theta}}(\theta) = H(\theta)$.  Thus  $H(\theta) = F_{M_{\theta}}(\theta) <  C_{E}(\theta)$.
 \end{enumerate}  

\begin{lemma}
Equality holds in (\ref{hlessce2}) if and only if the channel satisfies (\ref{wjwk0}) and we can find  a fixed unitary matrix $U = [u_{jk} ]$, such that the Kraus operators $E_j$ are related to the canonical Kraus operators $\Upsilon_k$ by
\begin{equation*}
E_j(\theta) = \sum_k u_{j k} \Upsilon_k(\theta).
\end{equation*}
\label{eq1ce}
\end{lemma}
\proof
From the proof of Lemma \ref{hlessce} we see that equality in (\ref{hlessce2}) is attainable only when the bound  given by the canonical Kraus operators $C_{\Upsilon}$ is attainable.  The bound  $C_{\Upsilon}$ is attainable if and only if the chanel satisfies (\ref{wjwk0}). It was shown \cite[p$372$]{chuang00} that if any two sets of Kraus operators lead to the same quantum channel they must be related in the following way:
\begin{equation}
E_j = \sum_k u_{j k} \Upsilon_k,
\label{EUF}
\end{equation}
where $U = [u_{jk} ]$ is a unitary matrix. When  $C_{\Upsilon}$ is attainable \cite{saromilb06},
\begin{equation*}
C_E = C_{\Upsilon} + 4 \sum_{jk} p_j | u_{jk}'|^2.
\end{equation*}
Thus for equality in (\ref{hlessce2}) we require further that $\sum_{jk} p_j | u_{jk}'|^2 = 0$. This is satisfied if and only if the entries $u_{jk}$ have no dependence on $\theta$ for all $k$ and $j, p_j \neq 0$, i.e.\ if and only if we can find a unitary matrix $U= [u_{jk} ]$ satisfying (\ref{EUF}) that has no dependence on $\theta$.

{\noindent{\bf{Remark 3}} }
We cannot use condition (\ref{eq:eupsrho}) generally to test for optimal POVMs. 
Condition  (\ref{eq:eupsrho}) is a necessary and sufficient condition for equality between the Fisher information and the SM bound. Since it is not generally possible to achieve equality between the Fisher information and the SM bound, condition (\ref{eq:eupsrho}) cannot be achieved for general models. Thus we cannot use it generally to test for optimal POVMs.

   \section{Multi-parameter channels}
\subsection{Introduction} \label{intro}
We shall look briefly at multi-parameter quantum estimation theory.  
The Fisher information for the parameter $\theta = (\theta^1, \dots \theta^m)$ is an $m \times m$ matrix, with entries
 \begin{eqnarray*}
 F_{M}(\theta)_{jk} &\equiv&  \int d\xi  p(\xi;\theta) \left( \frac{\partial \ln p(\xi;\theta)}{\partial \theta^j} \right)\left( \frac{\partial \ln p(\xi;\theta)}{\partial \theta^k} \right)    \\
&=&  \int d\xi \frac{1}{p(\xi;\theta)} \left( \frac{\partial p(\xi;\theta)}{\partial \theta^j} \right)\left( \frac{\partial p(\xi;\theta)}{\partial \theta^k} \right).
\label{eq:FI00}
\end{eqnarray*}
The Cram\'er--Rao inequality is a matrix inequality, which states that the covariance matrix of the estimator $t(x)$ is bounded below by the inverse of the Fisher information matrix, i.e. 
 \begin{equation}
 \mathrm{cov}_{\theta}[ t(x) ] \geq  F_{M}(\theta)^{-1}. 
 \label{eq:mqmcr}
 \end{equation}
The SLD quantum information is also an $m \times m$ matrix, with entries
\begin{equation*}
H(\theta)_{jk} = \Re \tr \left\{ \lambda^{(j)}  \rho(\theta)   \lambda^{(k)}  \right\},\\
\end{equation*}
where we define $\lambda^{(j)}$ as the SLD quantum score with respect to the parameter $\theta^j$, i.e.\
any self-adjoint solution of the equation,
 \begin{equation}
\frac{\partial \rho(\theta)}{\partial \theta^j} = \frac{1}{2}( \rho(\theta) \lambda^{(j)} +  \lambda^{(j)} \rho(\theta)).
\label{eq:sldeqn1}
\end{equation}
The inequality 
\begin{eqnarray}
F_M(\theta) \leq H(\theta),
\label{eq:multifh}
\end{eqnarray}
between the Fisher information and the SLD quantum information, holds but is not generally achievable in the multi-parameter case \cite{barndorffnielsengill00}. Putting (\ref{eq:mqmcr}) and (\ref{eq:multifh}) together we get the multi-parameter quantum Cram\'er--Rao inequality. This states that the covariance matrix of the estimator $t(x)$ is bounded below by the inverse of the SLD quantum information matrix, i.e.
\begin{equation}
 \mathrm{cov}_{\theta}[ t(x) ] \geq  H(\theta)^{-1}. 
 \label{qmcrineqm}
 \end{equation}
In contrast to the bound (\ref{eq:QCRineq}) in the one-parameter case, this bound cannot generally be achieved even asymptotically  \cite{barndorffnielsengill00}.  

Multi-parameter quasi-classical channels have output states of the form
 \begin{eqnarray}
\rho_{out}(\theta) = \sum_k p_k(\theta^1,\dots,\theta^m) | w_k \rangle \langle w_k |, \quad   \langle w_j | w_k \rangle = \delta_{jk},
\label{eq:quasicm}
\end{eqnarray}
where the eigenvectors $| w_k \rangle$ do not depend on $\theta$. The SLD quantum score with respect to the parameter $\theta^j$ is
 \begin{eqnarray*}
\lambda^{(j)} = \sum_k \frac{1}{p_k(\theta^1,\dots,\theta^m)}\frac{ \partial p_k(\theta^1,\dots,\theta^m)}{\partial \theta^j} | w_k \rangle \langle w_k |.
\label{eq:quasicms}
\end{eqnarray*}
Equality holds in (\ref{eq:multifh}) if and only if we can find a POVM $\{ M_m \}$ satisfying 
\begin{equation}
M_{m}^{1/2} \lambda^{(j)} \rho^{1/2} = \xi_m(\theta)  M_{m}^{1/2} \rho^{1/2}, \quad  \qquad \forall m,j.
\label{eq:eupsrho2}
\end{equation}
For a proof of (\ref{eq:eupsrho2}) see (1.25) of Ballester\cite{ballester05}. For quasi-classical channels of the form (\ref{eq:quasicm}), the POVM $\{ | w_k \rangle \langle w_k | \}$ satisfies (\ref{eq:eupsrho2})  and hence gives equality in (\ref{eq:multifh}) .

\subsection{The multi-parameter SM bound}
In this section we generalize the SM bound for general multi-parameter channels. We show that the multi-parameter SM bound is larger than or equal to the Fisher information matrix and is not achievable except in special situations. We define the multi-parameter SM bound as the matrix with entries,
    \begin{eqnarray*}
C_{\Upsilon}(\theta)_{jk} = 4 \sum_{l} \Re \tr \left\{ \Upsilon_{l}(\theta)^{(j)} \rho_{0}  \Upsilon_{l}(\theta)^{(k) \dagger}   \right\}, \qquad \Upsilon_{l}(\theta)^{(k)} = \frac{\partial }{\partial \theta^k}\Upsilon_{l}(\theta).
\end{eqnarray*}
    
    \begin{lemma}
The SLD quantum information is less than or equal to the SM bound for multi-parameter channels, i.e.
\begin{equation}
 H(\theta) \leq C_{\Upsilon}(\theta).
\label{mulithc}
\end{equation}
\label{lm:4}
\end{lemma}
\proof
Equation (\ref{mulithc}) is equivalent to 
\begin{equation}
 v^T   H(\theta)  v \leq  v^T  C_{\Upsilon}(\theta) v ,
\label{eq:cCc}
\end{equation}
for all $v \in \mathbb{R}^p$. To prove  (\ref{mulithc}) we choose suitable one-parameter channels and use Lemma \ref{eq:ineqcs}. For given $\theta$ and $ v$ in $\mathbb{R}^p$, consider the set of one-parameter channels
\begin{equation*}
 \rho_{0} \mapsto  \sum_{k}  \Upsilon_{k} (\theta + t  v ) \rho_{0} \Upsilon_{k}^{\dagger}(\theta + t v  ) ,\quad t \in \mathbb{R}.
\label{eq:multiproof}
\end{equation*}
Now,
\begin{eqnarray}
\frac{d}{dt}  \Upsilon_{k} (\theta + t  v  ) &=& \sum_l  \Upsilon_{k} (\theta)^{(l)} v^l + O(t),
\label{eq:multip1}\\
\tilde  \lambda(t)   &=&  \sum_l \tilde \lambda(\theta)^{(l)} v^l + O(t),\label{eq:multip3}
\end{eqnarray}
where $ v^l$ is the $l$th component of the vector $v$.  For singular states the SLD quantum  score $\lambda$ is not unique. In (\ref{eq:multip3}), we have chosen a specific SLD quantum score $\tilde  \lambda$ (see Appendix B). We prove equations (\ref{eq:multip1}) and (\ref{eq:multip3}) in Appendices C and D, respectively.
From Lemma \ref{eq:ineqcs} we know that $  H(t) \leq C_{\Upsilon}(t)$, i.e.
\begin{equation*}
 \tr \left\{ \tilde \lambda(t) \rho_{out}(\theta + t v) \tilde \lambda(t) \right\} \leq 4  \sum_{l=1}^{d} \tr \left\{ \frac{d}{d t}  \Upsilon_{l} (\theta + t v ) \rho_0  \frac{d}{d t}  \Upsilon_{l} (\theta + t  v  )^{\dagger} \right\}.
 \label{eq:34}
\end{equation*}
Using (\ref{eq:multip1}) and (\ref{eq:multip3}) and evaluating at $t=0$  gives
\begin{equation*}
\sum_{m,n} v^m v^n \tr \left\{ \tilde \lambda^{(m)} \rho_{out}(\theta) \tilde \lambda^{(n)} \right\} \leq 4 \sum_{m,n,l} 
v^m v^n \tr \left\{  \Upsilon_{l}(\theta)^{(m)} \rho_0  \Upsilon_{l}(\theta  )^{(n) \dagger} \right\}.
\label{eq:34b}
\end{equation*}
This is equivalent to (\ref{eq:cCc}). Since this holds for all $v$ in $\mathbb{R}^p$, we have (\ref{mulithc}). 

   \begin{lemma}
Equality holds in (\ref{mulithc}) for channels if and only if they satisfy
\begin{eqnarray}
\bigg\langle w_j^{(l)} \bigg| w_k \bigg\rangle = 0, \quad \forall j,k,l,   \quad \mathrm{with} \quad   p_j,p_k > 0, \quad    \bigg| w_j^{(l)} \bigg\rangle = \frac{\partial}{\partial \theta^l}   | w_j \rangle.
\label{eq:unitcond3}
\end{eqnarray}
\label{eq:equalitym}
\end{lemma}
\proof
Equality in (\ref{mulithc}) is equivalent to 
\begin{equation}
 v^T   H(\theta)  v =  v^T  C_{\Upsilon}(\theta) v ,
\label{eq:cCc2}
\end{equation}
for all $v \in \mathbb{R}^p$.  From the proof of the previous lemma we see that for (\ref{eq:cCc2}) to be satisfied for all $v \in \mathbb{R}^p$, we require that, for one-parameter channels of the form 
\begin{equation*}
 \rho_{0} \mapsto  \sum_{k}  \Upsilon_{k} (\theta + t  v  ) \rho_{0} \Upsilon_{k}^{\dagger}(\theta + t v  ) ,\quad t \in \mathbb{R},
 \label{eq:ello1}
\end{equation*}
for given $\theta$ and $v \in \mathbb{R}^p$, we have $H(t) |_{t=0} =  C_{\Upsilon}(t) |_{t=0}$. From Lemma \ref{lemmylem} this is possible if and only if the channel satisfies (\ref{wjwk0}) at the point $t=0$. This condition is equal to
\begin{equation*}
\left. \left( \frac{d}{dt} \langle w_j | \right) w_k \rangle\right|_{t=0} = 0, \quad \forall j,k,  \quad \mathrm{with} \quad   p_j,p_k > 0. 
\end{equation*}
This condition can be rewritten as (Appendix D) 
  \begin{eqnarray}
\sum_{l=1}^{m} v^l \bigg\langle w_j^{(l)} \bigg| w_k \bigg\rangle = 0 \quad \forall j,k,  \quad \mathrm{with} \quad   p_j,p_k > 0.
\label{eq:unitcond2}
\end{eqnarray}
Condition (\ref{eq:unitcond2}) holds for all $v$ if and only if (\ref{eq:unitcond3}) is satisfied.

\begin{lemma}
For channels for which $p_j(\theta) > 0$ for all $j$ and $\theta$, equality holds in (\ref{mulithc}) if and only if the channel is quasi-classical.
\end{lemma}
\proof
This follows from (\ref{eq:unitcond3}) and the same analysis as Lemma \ref{lemc}.

  \begin{lemma}
For unitary channels, equality holds in (\ref{mulithc}) if and only if
\begin{eqnarray*}
\tr \left\{ U(\theta) \rho_{0} \frac{\partial U(\theta)}{\partial \theta^l}^{\dagger}  \right\} = 0, \quad \forall l.
\label{eq:unitcond}
\end{eqnarray*}
\end{lemma}
\proof
This follows from (\ref{eq:unitcond3}) and and the same analysis as Lemma \ref{lemd}.

{\noindent{\bf{Example 2}} }
There exist channels which are neither quasi-classical or unitary for which equality holds in  (\ref{mulithc}). The channel with the following output states satisfies (\ref{eq:unitcond3}) and hence achieves equality in (\ref{mulithc}):
\begin{equation*}
\rho_{out}(\theta) = f(\theta)^2 | v_1(\theta) \rangle \langle v_1(\theta)  | + ( 1 - f(\theta)^2 ) | v_2(\theta)  \rangle \langle v_2 (\theta)|,
\end{equation*}
where $f(\theta)$ and $g(\theta)$ are real functions of $\theta$ with $\quad 0 \leq f(\theta), g(\theta) \leq 1$  and 
\begin{equation*}
| v_1(\theta)  \rangle = ( g(\theta), \sqrt{ 1 - g(\theta)^2}, 0 )^T, \quad  | v_2(\theta)  \rangle = ( 0, 0, 1 )^T.
\end{equation*}

\begin{theorem}
For multi-parameter channels, the SM bound is an upper bound on the Fisher information, i.e.
\begin{equation}
F_{M}(\theta) \leq C_{\Upsilon}(\theta).
\label{mulithc1}
\end{equation}
\end{theorem}
\proof
This follows from  (\ref{eq:multifh}) and (\ref{mulithc}). 

 \begin{lemma}
The only channels for which equality is attainable in (\ref{mulithc1})  are channels which satisfy (\ref{eq:unitcond3}) and for which there exists a POVM satisfying (\ref{eq:eupsrho2}).
\end{lemma}

\proof
For equality in  (\ref{mulithc1}) we require equality in both (\ref{eq:multifh}) and (\ref{mulithc}). Equality is attainable in (\ref{mulithc}) only for channels satisfying (\ref{eq:unitcond3}).  Equality is attainable in (\ref{eq:multifh}) if and only if there exists a POVM satisfying (\ref{eq:eupsrho2}).

\begin{lemma}
\begin{eqnarray}
H(\theta) \leq C_{E}(\theta), \quad \forall E.
\label{multihlessce2}
\end{eqnarray}
\label{multihlessce}
  \end{lemma}
  \proof
This follows from  (\ref{hlessce2}) and the analysis used in the proof of Lemma \ref{lm:4} with $\Upsilon_k$ replaced by $E_k$.

\begin{lemma}
Equality holds in (\ref{multihlessce2}) if and only if the channel satisfies (\ref{eq:unitcond3}) and we can find a fixed unitary matrix $U = [ u_{jk} ]$ such that the Kraus operators $E_j$ are related to the canonical Kraus operators $\Upsilon_k$ by
 \begin{equation}
E_j(\theta) = \sum_k u_{j k} \Upsilon_k(\theta).
\label{EUF2}
\end{equation}
\end{lemma}
\proof
For given $\theta$ and $v$ in $\mathbb{R}^p$, consider the set of channels
\begin{equation}
 \rho_{0} \mapsto  \sum_{k}  \Upsilon_{k} (\theta + t v ) \rho_{0} \Upsilon_{k}^{\dagger}(\theta + t  v  ) ,\quad t \in \mathbb{R}.
\label{eq:multiproof}
\end{equation}
For equality in (\ref{multihlessce2}) we require that for all $v$ we have $\left. H(t) \right|_{t=0} = \left. C_E (t)\right|_{t=0}$.
From Lemma \ref{eq1ce} this is satisfied if and only if the channel satisfies (\ref{wjwk0}) at $t=0$ and the Kraus operators $E_j$ are related to the canonical Kraus operators $\Upsilon_k$ by
\begin{equation*}
\left. E_j(\theta+tv) \right|_{t=0} =  \sum_k u_{j k}(\theta+tv) \Upsilon_k(\theta+tv)\left. \right|_{t=0},
\end{equation*}
where
\begin{equation}
 \sum_{jk} p_j \left| \left. \frac{d u_{jk}}{d t}\right|_{t=0} \right|^2= 0.
\label{unitzar}
\end{equation}
From the proof of Lemma \ref{eq:equalitym} we see that for channels of the form (\ref{eq:multiproof}),  satisfying  (\ref{wjwk0}) at $t=0$ is equivalent to satisfying (\ref{eq:unitcond3}).
We can re-write (\ref{unitzar}) as
\begin{equation*}
 \sum_{jk}  p_j  \left|  \sum_l \frac{\partial u_{jk}}{\partial \theta^l} v^l \right|^2 =  0.
\end{equation*}
This is satisfied for all $v$ if and only if the entries $u_{jk}$ have no dependence on $\theta$ for all $k$ and $j, p_j \neq 0$, i.e.\ if and only if we can find a unitary matrix $U= [u_{jk} ]$ satisfying (\ref{EUF2}) that has no dependence on $\theta$.

\begin{lemma}
\begin{eqnarray}
  F_{M}(\theta) \leq C_{E}(\theta), \quad \forall E.
  \label{multiflessce}
\end{eqnarray}
  \end{lemma}
  \proof
 This follows from (\ref{eq:multifh}) and (\ref{multihlessce2}).
 
 \section{Conclusion}
We have clarified the relation between the SM bound and the SLD quantum information and the equality conditions. In doing so the question of attainability of the Sarovar and Milburn bound for one-parameter channels has been settled. The attainability conditions of the SM bound (\ref{eq:eupsrho}) cannot be used generally to test for optimal POVMs. We have extended the inequality between the SM bound and the SLD quantum information for multi-parameter channels. Consequently, the SM bound is greater than or equal to the Fisher information for multi-parameter channels. 

\ack
I am very grateful to Peter Jupp for informative discussions on geometry and numerous helpful comments.
This work was supported by the EPSRC. 

\appendix

\section*{Appendix A}
\setcounter{section}{1}
Here it is shown that $C_{\Upsilon}(\theta)$ can be written as (\ref{eq:samr}).
When $p_j = 0$,
 \begin{eqnarray*}
 \Upsilon_{j}| \psi_{0} \rangle &=& \sqrt{p_{j}} | w_{j}  \rangle = 0,\\
\Upsilon_{j}' | \psi_{0} \rangle &=&  0, \\
  \tr \{  \Upsilon_{j}^{'}  \rho_{0} \Upsilon_{j}^{\dagger '}  \} &=& \langle \psi_{0} | \Upsilon_{j}^{\dagger '} \Upsilon_{j}' | \psi_{0}  \rangle = 0.
  \end{eqnarray*} 
 When $p_j \neq 0$,
\begin{eqnarray*}
 \Upsilon_{j}| \psi_{0} \rangle &=& \sqrt{p_{j}} | w_{j}  \rangle,\\
\Upsilon_{j}' | \psi_{0} \rangle &=& \frac{p_{j}'}{2 \sqrt{p_{j}}}  | w_{j}  \rangle +   \sqrt{p_{j}} | w_{j} ' \rangle .\\
\end{eqnarray*}
Then
\begin{eqnarray*}
\langle \psi_{0} | \Upsilon_{j}^{\dagger '} \Upsilon_{j}' | \psi_{0}  \rangle &=& \left(\frac{p_{j}'}{2 \sqrt{p_{j}}}  \langle w_{j} | +   \sqrt{p_{j}} \langle w_{j} '| \right) \left(\frac{p_{j}'}{2 \sqrt{p_{j}}}  | w_{j}  \rangle +   \sqrt{p_{j}} | w_{j} ' \rangle \right),\\
&=& \frac{p_{j}'^2}{4p_{j}} + \frac{p_{j}'}{2} \left( \langle w_{j}' | w_{j}  \rangle + \langle w_{j} | w_{j}'  \rangle \right) + p_{j} \langle w_{j}' | w_{j}'  \rangle.
\end{eqnarray*}
The previous line simplifies, because
\begin{eqnarray}
\langle w_{j}' | w_{j}  \rangle + \langle w_{j} | w_{j}'  \rangle = \frac{\partial}{\partial \theta}  \tr \{\rho_{j} \}  = 0 , \qquad  \rho_{j} =  | w_{j} \rangle  \langle w_{j} |.
\label{eq:diffjj}
\end{eqnarray}
Thus we have 
\begin{eqnarray*}
C_{\Upsilon}(\theta)&=& 4 \sum_{j, p_j \neq 0}\left( \frac{p_{j}'^2}{4p_{j}} + p_{j}\langle w_{j}' | w_{j}'  \rangle \right). 
\end{eqnarray*}
If we insert the identity $I = \sum_{k=1}^{d} | w_{k}  \rangle  \langle   w_{k} |$ into $\langle w_{j}' | w_{j}'  \rangle $ we get
\begin{eqnarray*}
C_{\Upsilon}(\theta) &=& \sum_{j, p_j \neq 0}  \frac{p_{j}'^2}{p_{j}} + \sum_{j, k, p_j \neq 0}  4 p_{j}\langle w_{j}' | w_{k}  \rangle  \langle   w_{k} |  w_{j}'  \rangle, \\
&=&  \sum_{j, p_j \neq 0}  \frac{p_{j}'^2}{p_{j}} +  \sum_{j, k, p_j \neq 0}  4 p_{j} |\langle w_{j}' | w_{k}  \rangle |^2.
\end{eqnarray*}
This can be re-written, since
\begin{eqnarray}
\langle w_{j} | w_{k} \rangle &=& \delta_{jk}, \nonumber \\
\frac{\partial}{\partial \theta} \langle w_{j} | w_{k}  \rangle &=& \langle w_{j}' | w_{k}  \rangle + \langle w_{j} | w_{k}'  \rangle = 0,\nonumber \\
\langle w_{j}' | w_{k}  \rangle &=& - \langle w_{j} | w_{k}'  \rangle, \label{eq:wjwk} \nonumber \\
|\langle w_{j}' | w_{k}  \rangle |^2 &=& \langle w_{j}' | w_{k}  \rangle  \langle w_{k} | w_{j}'  \rangle \nonumber \\
&=& (- \langle w_{j} | w_{k}'  \rangle) (- \langle w_{k}' | w_{j}  \rangle) =  |\langle w_{k}' | w_{j}  \rangle |^2. \label{eq:diffwjwk}
\end{eqnarray}
Now,
\begin{eqnarray*}
\sum_{j, k, p_j \neq 0}  4 p_{j} |\langle w_{j}' | w_{k}  \rangle |^2 &=& \sum_{j < k, p_j \neq 0}  4 p_{j} |\langle w_{j}' | w_{k}  \rangle |^2 + \sum_{ k < j , p_j \neq 0}  4 p_{j} |\langle w_{j}' | w_{k}  \rangle |^2  \\
&+& \sum_{j = k, p_j \neq 0}  4 p_{j} |\langle w_{j}' | w_{k}  \rangle |^2.
\end{eqnarray*}
Swapping the indices $j$ and $k$ in the second term and using (\ref{eq:diffwjwk}) simplifies this further to 
\begin{eqnarray*}
 \sum_{j, k, p_j \neq 0}  4 p_{j} |\langle w_{j}' | w_{k}  \rangle |^2 = \sum_{j < k, p_j + p_k \neq 0}  4( p_{j} + p_k ) |\langle w_{j}' | w_{k}  \rangle |^2 + \sum_{j , p_j \neq 0}  4 p_{j} |\langle w_{j}' | w_{j}  \rangle |^2.
\end{eqnarray*}
We can rewrite the SM bound as
\begin{eqnarray}
C_{\Upsilon}(\theta) = \nonumber \\ 
\sum_{j,  p_j \neq 0}  \frac{p_{j}'^2}{p_{j}}    + \sum_{j < k,  p_j+p_k > 0}  4 ( p_{j} + p_{k} ) |\langle w_{j}' | w_{k}  \rangle |^2  +   4\sum_{ p_k \neq 0}  p_{k} |\langle w_{k}' | w_{k}  \rangle |^2. 
\end{eqnarray}

\appendix
\section*{Appendix B}
\setcounter{section}{2}
In this section it is shown that for output states of the form (\ref{eq:outstate}) the SLD quantum information is of the form  (\ref{eq:Hsldr}). The SLD is defined as any self-adjoint solution $\lambda$ of the matrix equation 
\begin{equation}
\frac{d \rho(\theta)}{d\theta} = \frac{1}{2}\left\{\rho(\theta)\lambda(\theta)+\lambda(\theta)\rho(\theta)\right\}.
\label{sld:def}
 \end{equation}
The SLD quantum information is defined as
\begin{equation*}
H(\theta) = \tr \{ \rho \lambda^2 \} .
\label{Hsld:def2}
\end{equation*}
Substituting (\ref{eq:outstate}) into (\ref{sld:def}), we get 
\begin{eqnarray}
\sum_{i=1} \left\{ p_{i}' | w_i \rangle \langle  w_i | + p_{i} (  | w_{i}'  \rangle \langle w_i | + | w_i  \rangle \langle w_{i} ' | ) \right\} \nonumber \\ =   \frac{1}{2} \left( \sum_{l} p_{l}| w_l  \rangle \langle w_l | \lambda + \lambda \sum_{m}  p_{m} | w_m  \rangle \langle w_m | \right) . 
\label{sld:def2}
 \end{eqnarray}
From (\ref{sld:def2}) we calculate the components of the SLD. First, we consider the diagonal elements $\lambda_{jj}$.  Pre-multiply (\ref{sld:def2}) by $\langle w_j |$
 and post-multiply it by $| w_j \rangle$.  On the left hand side we get
 \begin{equation*} 
 p_{j}' +   p_{j} ( \langle w_j  | w_{j}'  \rangle + \langle w_{j}'  | w_{j}  \rangle ) = p_{j}' 
 \end{equation*}
by (\ref{eq:diffjj}). On the right hand side we get 
 \begin{equation*} 
p_{j} \langle w_j | \lambda | w_j \rangle.
  \end{equation*}
Hence, provided that $p_j \neq 0$ 
 \begin{equation*} 
\lambda_{jj} = \frac{p_{j}'}{p_j}. 
  \end{equation*}
The diagonal elements   $\lambda_{jj}$  are not defined when $p_j = 0$. In this case, we choose a particular solution of $\lambda$ for which $\lambda_{jj} =0$.  Next, we look at the off-diagonal components $\lambda_{jk}$ of the SLD. We pre-multiply (\ref{sld:def2})  by $\langle w_j |$ and post-multiply it by $| w_k \rangle$.  On the left hand side we get
 \begin{equation*} 
0 + p_{k}  \langle w_j | w_{k}' \rangle + p_{j}  \langle w_{j}' | w_k \rangle = (p_{j} - p_{k}) \langle w_{j}' | w_k \rangle
  \end{equation*}
by (\ref{eq:wjwk}). On the right hand side we get
\begin{equation*} 
\frac{1}{2} (p_{j} + p_{k})  \langle w_j | \lambda | w_k \rangle.
  \end{equation*}
So, provided that $p_j + p_k > 0$, we get 
\begin{equation*} 
\lambda_{jk} = \frac{ 2 (p_{j} - p_{k}) \langle w_{j}' | w_k \rangle }{p_{j} + p_{k}}.
  \end{equation*}
The entries  $\lambda_{jk}$ are not defined when $p_j + p_k = 0$. Again we choose a particular solution of $\lambda$ for which $\lambda_{jk} =0$, when $p_j + p_k = 0$.   We define a particular solution  of the SLD as \begin{eqnarray}
\tilde \lambda =  \sum_{k, p_k \neq 0} \frac{p_{k} '}{p_{k}} | w_k \rangle \langle w_k | +  \nonumber  \\
 \sum_{   j < k, p_j + p_k > 0 } 2 \frac{ ( p_{j} - p_{k})}{p_{j} + p_{k}}\left( \langle w_{j}' | w_{k} \rangle | w_j \rangle \langle  w_k | + \langle w_{k} | w_{j}' \rangle  | w_k \rangle \langle  w_j | \right).
\label{eq:Lsldr}
\end{eqnarray}
  From this we get the required result.

\appendix
\section*{Appendix C}
\setcounter{section}{3}
Here, we prove that 
\begin{equation}
\frac{d }{d t}  \Upsilon_{k} (\theta + t v ) = \sum_l  \Upsilon_{k} (\theta)^{(l)} v^l + O(t). 
\label{eq:multip1x}
\end{equation}
Let us introduce the parameter $\phi(t) = \theta + t v $,  with components $\phi^l = \theta^l + t  v ^l $. Using the chain rule to differentiate $\Upsilon_{k} (\phi(t) )$, we get
\begin{equation}
\frac{d}{d t}  \Upsilon_{k} (\phi(t)) = \sum_l  \frac{ \partial \Upsilon_{k} (\phi)}{\partial \phi^l} \frac{ \partial \phi^l}{\partial t} .
\label{eq:appca8}
\end{equation}
Now,
\begin{eqnarray*}
  \frac{ \partial \Upsilon_{k} (\phi)}{\partial \phi^l}  &=&  \left. \frac{ \partial \Upsilon_{k} (\phi)}{\partial \phi^l}  \right|_{t=0}+ O(t) = \frac{ \partial \Upsilon_{k} (\theta)}{\partial \theta^l} + O(t) ,\\
  \frac{ \partial \phi^l}{\partial t}  &=& v^l.
  \end{eqnarray*}
Substituting these back into (\ref{eq:appca8}) gives (\ref{eq:multip1x}).

\appendix
\section*{Appendix D}
\setcounter{section}{4}
In this appendix we prove that 
 \begin{equation}
\tilde \lambda(t) = \sum_{l=1}^m \tilde \lambda^{(l)} v^l + O(t). 
\label{eq:lamtilde}
 \end{equation}
 Now
 \begin{eqnarray*}
p_k(t) &=& p_k (\theta) + O(t),\label{O1} \\
 \frac{d p_k(t)}{d t } &=& \sum_l  p_k^{(l)}  v^{l} + O(t) , \quad p_k^{(l)} = \frac{\partial p_k(\theta)}{ \partial \theta^{l}}  \label{O2}\\
 \frac{d | w_k(t) \rangle }{d t } &=&  \sum_l  \bigg| w_k^{(l)} \bigg\rangle  v^l + O(t), \quad \bigg| w_k^{(l)} \bigg\rangle  = \frac{\partial | w_k(\theta) \rangle}{ \partial \theta^l}.
  \label{O3}
 \end{eqnarray*}
 Substituting the above equations into (\ref{eq:Lsldr}) gives

 \begin{eqnarray*}
\tilde \lambda(t) =  \sum_{k, p_k \neq 0}  \frac{ \sum_l  p_k^{(l)}  v^{l} + O(t)}{p_k (\theta) + O(t)} | w_k (t) \rangle \langle w_k (t) |   \nonumber  \\
 + \sum_{  j < k, p_j + p_k > 0} 2 \frac{ ( p_j (\theta)  - p_k (\theta) + O(t))}{p_j (\theta) +  p_k (\theta) + V(t)} A_{jk} , \nonumber\\
 A_{jk} =  \left( \left( \sum_l v^l  \bigg\langle w_j^{(l)} \bigg| w_{k} \bigg\rangle   + O(t)\right) \bigg| w_j \bigg\rangle \bigg\langle w_k \bigg| + \left( \sum_l v^l  \bigg\langle w_k \bigg|  w_j^{(l)} \bigg\rangle   + O(t)\right)  \bigg| w_k \bigg\rangle \bigg\langle  w_j \bigg| \right).\nonumber
\label{eq:Lsldr2}
\end{eqnarray*}
Thus $\tilde \lambda(t)$ has the form (\ref{eq:lamtilde}).

\end{document}